\documentclass[preprint]{elsarticle}

\usepackage{lineno,hyperref}

\usepackage{algorithm}
\usepackage{algpseudocode}

\journal{Journal of Parallel and Distributed Computing}

\begin{document}

\begin{frontmatter}

\title{Parallel Louvain Community Detection Optimized for GPUs}

\author{Richard Forster}
\address{}

\author[address1]{CERN}
\ead[url]{www.home.cern}

\author[address2]{Eotvos Lorand University\corref{mycorrespondingauthor}}
\cortext[mycorrespondingauthor]{Corresponding author}
\ead{forceuse@inf.elte.hu}
\ead[url]{www.inf.elte.hu}

\author[address2]{Wigner RCP}
\ead[url]{www.wigner.mta.hu}

\address[address1]{Geneva, Switzerland}
\address[address2]{Budapest, Hungary}

\begin{abstract}
Community detection now is an important operation in numerous graph based applications. It is used to reveal groups that exist within real world networks without imposing prior size or cardinality constraints on the set of communities. Despite its potential, the support for parallel computers is rather limited. The cause is largely the irregularity of the algorithm and the underlying heuristics imply a sequential nature. In this paper a GPU based parallelized version of the Louvain method is presented. The Louvain method is a multi-phase, iterative heuristic for modularity optimization. It was originally developed by Blondel et al. (2008), the method has become increasingly popular owing to its ability to detect high modularity community partitions in a fast and memory-efficient manner. The parallel heuristics used, were first introduced by Hao Lu et al. (2015). As the Louvain method is inherently sequential, it limits the possibility of scalable usage. Thanks to the proposed parallel heuristics, I observe how this method can behave on GPUs. For evaluation I implemented the heuristics using CUDA on a GeForce GTX 980 GPU and for testing I’ve used organization landscapes from the CERN developed Collaboration Spotting project that involves patents and publications to visualize the connections in technologies among its collaborators. Compared to the parallel Louvain implementation running on 8 threads on the same machine that has the used GPU, the CUDA implementation is able to produce community outputs comparable to the CPU generated results, while providing absolute speedups of up to 30 using the GeForce GTX 980 consumer grade GPU.
\end{abstract}

\begin{keyword}
CUDA\sep GPU\sep Louvain\sep Modularity\sep Community Detection\sep parallel
\end{keyword}

\end{frontmatter}


\section{Introduction}
Applications working with huge datasets requires optimizations to be able to utilize the available system resources efficiently. To leverage the power of the special architectures and their parallel capabilities additional heuristics might be needed, that can drive the otherwise sequential algorithm to run on multiple threads or even on a number of machines at the same time. Community detection is such an application, that is by default a sequential problem involving multiple phases and iterations to produce the desired community partition.

For community detection the Louvain algorithm will be explored and it's implementation will be detailed, that involves parallel heuristics, that are required to be able to achieve the aforementioned efficiency. Furthermore the goal of this paper is to show the algorithm running on a GPU, thus delivering much higher performance, than what is possible on the CPU. The algorithm will be explained in details, followed by implementation techniques that involves the usage of the CUDA platform, that is a GPU dedicated programming toolkit, that can be used to do generic computational applications running on NVIDIA GPUs.

The performance of the algorithm will be explored in both experimental and theoretical ways, as the networks used for illustration couldn't all fit on the test system. The experimental study shows a $23$ times speed up over the CPU based parallel implementation, while the theoretical results are pointing to an even higher $31$ times faster runtime.

\section{Problem statement and notation}\label{prob}
Let $G(V,E,\omega)$ be an undirected weighted graph, with $V$ representing the set of vertices, $E$ the set of edges and $\omega$ a weighting function that assigns a positive weight to every edge from $E$. If the input graph is unweighted, then the weight of the edges is considered to be $0$. The graph is allowed to have loops, so edges like $(i,i)$ are valid, while multiple edges between the same nodes should not be present. The following will be the adjacency list of vertex $i$: $\Gamma(i)={j|(i,j)\in E}$. Let $\delta_i$ denote the weighted degree of vertex $i$, such as $\delta_i=\sum_{j\in\Gamma(i)}\omega(i,j)$. Let $N$ denote the number of vertices in graph $G$, $M$ the number of edges, and $W$ the sum of all edge weights, such as $M=\frac{1}{2}\sum_{i\in V}\delta_i$.
By computing the communities, the vertex set $V$ will be partitioned into an arbitrary number of disjoint subsets, each with size $n$, where $0<n\leq N$. $C(i)$ will denote the community containing vertex $i$. $E_{i\rightarrow C}$ is the set of all edges connecting vertex $i$ into community $C$. Consequently let $e_{i\rightarrow C}$ hold the sum of the edge weights in $E_{i\rightarrow C}$.

\begin{equation}
\label{eq1}
e_{i\rightarrow C}=\sum\limits_{(i,j)\in E_{i\rightarrow C}}\omega(i,j)
\end{equation}

The sum of all the vertices in community C shall be denoted by $deg_C$, which will represent the degree of the whole community.

\begin{equation}
\label{eq2}
deg_C=\sum\limits_{i\in C}\delta_i
\end{equation}

\subsection{Modularity}\label{modularity}
Let $S={C_1,C_2,...,C_k}$ be the set of every community in a given partitioning of $V$, where $1\leq k \leq N$. Modularity $Q$ of partitioning $S$ is given by the following \cite{findcom}:

\begin{equation}
\label{eq3}
Q=\frac{1}{2W}\sum\limits_{i\in V}e_{i\rightarrow C(i)}-\sum\limits_{C\in S}(\frac{deg_C}{2W}\cdot{}\frac{deg_C}{2W})
\end{equation}

Modularity is widely used in different community detection algorithms, while it also has issues, such as the resolution limit \cite{comdet}\cite{rescom}. The definition itself also has multiple variants \cite{rescom}\cite{modgraph}\cite{comdetres}. However, the version defined in Eq. \ref{eq3} is still the more commonly used. This definition is used in the Louvain method \cite{louvainalg}.

\subsection{Community detection}\label{comdetection}
On a given $G(V,E,\omega)$ the goal of community detection is to compute partitioning $S$ of communities that produces the highest modularity. This problem has been shown to be NP-Complete \cite{clustering}. The main difference between this problem and other graph partitioning is that, the number and size of the clusters and their distribution are known at the beginning. \cite{graphpart} For community detection, both quantities are unknown for the computation.

\section{The Louvain algorithm}\label{louvain}
In 2008 Blondel et al. presented an algorithm for community detection\cite{louvainalg}. The Louvain method, is a multi-phase, iterative, greedy algorithm used to produce the community hierarchy. The main idea is the following: the algorithm has multiple phases, each phase with multiple iterations that is running until a convergence criteria is met. At the beginning of the first phase each vertex is going to be assigned to a separate community. Going on, the algorithm progresses from one iteration to another until the modularity gain becomes lower than a predefined threshold. With every iteration, the algorithm checks all the vertices in an arbitrary, but predefined order. For every vertex $i$, all its neighboring communities (where the neighbors can be found) are explored and the modularity gain is computed for every possible community transfer. Once this calculation is done, the neighboring community yielding the highest modularity will be selected and vertex $i$ will be moved there. If no suitable community can be found the vertex stays in its own group. An iteration ends once all vertices are checked. As a result the modularity is a monotonically increasing function, that spreads across the multiple iterations of a phase. When the algorithm converges within a phase, it moves to the next one by reducing all vertices of a community to a single "meta-vertex"\cite{parlouvain}. The edge on such vertex can be a loop, in which case the weight will be the sum of the weights of all the edges that are having start and end vertices within that same community. If the edge is pointing into another community, the weight will be the sum of weights of all the edges between the corresponding two communities. The result will be graph $G'(V',E',\omega')$, which then becomes the input for the consecutive phase. Multiple phases are processed until the modularity converges. At any given iteration, let $\Delta Q_{i\rightarrow C(j)}$ represent the modularity gain resulting from moving vertex $i$ from its current community $C(i)$ to a neighboring $C(j)$. This is given by:

\begin{equation}
\label{eq4}
\Delta Q_{i\rightarrow C(j)}=\frac{e_{i\rightarrow C(j)}}{W}+\frac{2\cdot{}\delta_i\cdot{}mod_{C(i)/{i}}-2\cdot{}\delta_i\cdot{}mod_{C(j)}}{(2W)^2}
\end{equation}

The new community assignment for vertex $i$ will be determined as follows. For $j\in \Gamma(i)\cup{i}$:

\begin{equation}
\label{eq5}
C(i)=arg \max\limits_C(j)\Delta Q_{i\rightarrow C(j)}
\end{equation}

 Effectively the iterations and phases run forever, but because the modularity is monotonically increasing, it is guaranteed to terminate at some point. Generally the method needs only tens of iterations and fewer phases to terminate on real world datasets.

\subsection{Parallel heuristics}\label{heuristics}
The challenges to parallelize the Louvain method were explored in \cite{parlouvain}. To solve those issues multiple heuristics were introduced, that can be used to leverage the performance of the parallel systems in a basically sequential algorithm. From the proposed heuristics two is going to be detailed in this paper. Lets assume the communities at any given stage are labeled numerically. The notation $l(C)$ will return the label of community $C$.

\subsubsection{Singlet minimum label heuristic}\label{singletmin}
In the parallel algorithm, if at any given iteration vertex $i$ which is in a community by itself ($C(i)={i}$, singlet community \cite{parlouvain}), in hope for modularity gain might decide to move into another community, that holds only vertex $j$. This transition will only be applied if $l(C(j) < l(C(i)))$.

\subsubsection{Generalized minimum label heuristic}\label{genmin}
In the parallel algorithm, if at any given iteration the vertex $i$ has multiple neighboring communities providing modularity gains, the community with the minimum label will be selected. Swap situations might occur, when two vertices are transitioning into the other's community in the same iteration. This can delay the convergence, but can never lead to nontermination as the minimum required modularity gain threshold will guarantee a successful termination.

\newtheorem{thm}{Theorem 1}

\begin{thm}
	The memory requirement is not increasing between phases.
\end{thm}\label{thm1}

\newproof{pot1}{Proof of Theorem 1}

\begin{pot1}
	Because of the generalized minimum label heuristic, the nodes are always moving to the communities with lower labels. So as the computation is progressing, the clusters are either merging into other clusters with lower labels, or nodes are keep filling up the cluster from those, that have higher labels. This means the nodes are moving among the clusters, that have been defined at the beginning with the first initialization, thus not creating more clusters, which leads to keeping the memory footprint the same.
\end{pot1}

\section{Compute Unified Device Architecture}\label{CUDA}
\begin{figure}[!t]
	\centering
	\includegraphics[width=2.5in]{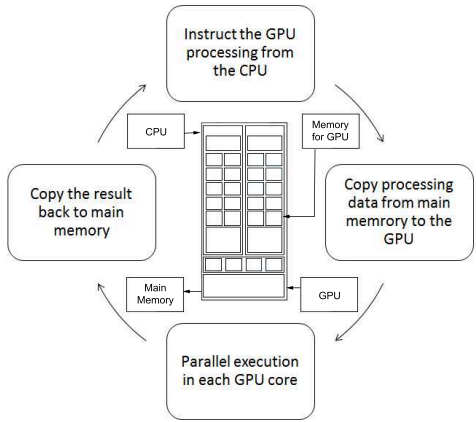}
	\caption{CUDA Processing Flow}
	\label{flow}
\end{figure}

Thanks to the modern GPUs, very big datasets can be efficiently computed in parallel \cite{cudaposter}\cite{cudaarticle}\cite{cudaarticle2}. This is supported by the underlying hardware architecture that allows us to create general purpose algorithms running on the graphical hardware. There is no need to introduce any middle abstractions to be able to use these processors as they have evolved into a more general processing unit (Figure \ref{flow}). The compute unified device architecture (CUDA) divides the GPUs into smaller logical parts to have a deeper understanding of the platform \cite{cuda}.
With the current device architecture the GPUs are working like coprocessors in the system, the instructions are issued through the CPU. In earlier generations the required data had to be copied manually over to the device memory. Furthermore the different parts in memory (global, shared, constant) used a different address space.

With the second generation of Compute Capability devices it has became possible to issue direct memory transactions thanks to the Unified Virtual Addressing \cite{fermi}. This has made it possible to use pointers and memory allocations not only on the host side, but on the device as well and also making it possible to use C/C++ like pointers.

\subsection{Memory access strategies}\label{memory}
For the early CUDA capable generation of GPUs with Compute Capability (CC) 1.x, it is important to use the right access patterns when operating on the device's global memory. In such a case where the same value is required from thousands of threads, then the operations will have to be serialized on these GPUs. The newer GPUs since CC 2.0 also have caching functionality that greatly improves the performance of such applications, increasing the effective throughput of the memory. To fetch and store data in the most optimal way, the map access strategy should be used, where all threads will manipulate their own individual values, more specifically thread $n$ will access the $nth$ position of the input or output array. The values about memory transactions presented in this subsection are taken from \cite{cuda}.

If the data stored in memory can be accessed in a sequential order and those values are aligned to a multitude of $128$ byte address, then the most optimal memory utilization will be achieved even on devices with CC 1.x (Table \ref{aligned}). The GPU can issue $32$, $64$ or $128$ bytes transactions based on the actual usage of the hardware.

\begin{table}[!th]
	\caption{Aligned and sequential memory access}
	\label{aligned}
	\centering
	\begin{tabular}{|c|c|c|c|}\hline
		CC & 1.0 and 1.1 & 1.2 and 1.3 & 2.x and 3.0\\ \hline
		Transactions & \multicolumn{2}{|c|}{Uncached} & Cached\\\cline{2-4}
		& 1 x 64B at 128 & 1 x 64B at 128 & 1 x 128B at 128\\
		
		& 1 x 64B at 192 & 1 x 64B at 192 &\\ \hline
	\end{tabular}
\end{table}

If the data is in a non-sequential order (Table \ref{nonsequential}), then the GPU will need additional memory transactions to process all the available values. We have to mention here, by using non-sequential ordering it is possible the transactions will fetch not only the actually required data, but also additional values that are stored on consecutive memory addresses. This can be quite a wasteful approach.

\begin{table}[!th]
	\caption{Aligned and non-sequential memory access}
	\label{nonsequential}
	\centering
	\begin{tabular}{|c|c|c|c|}\hline
		CC & 1.0 and 1.1 & 1.2 and 1.3 & 2.x and 3.0\\ \hline
		Transactions & \multicolumn{2}{|c|}{Uncached} & Cached\\\cline{2-4}
		& 8 x 32B at 128 & 1 x 64B at 128 & 1 x 128B at 128\\
		& 8 x 32B at 160 & 1 x 64B at 192 &\\
		& 8 x 32B at 192 & &\\
		& 8 x 32B at 224 & &\\ \hline
	\end{tabular}
\end{table}

If the data is completely misaligned (Table \ref{misaligned}), then the hardware will be forced to invoke more smaller transactions to reach all the values present for processing. This case can be found difficult even for devices with caching involved.

\begin{table}[!th]
	\caption{Misaligned and sequential memory access}
	\label{misaligned}
	\centering
	\begin{tabular}{|c|c|c|c|}\hline
		CC & 1.0 and 1.1 & 1.2 and 1.3 & 2.x and 3.0\\ \hline
		Transactions & \multicolumn{2}{|c|}{Uncached} & Cached\\\cline{2-4}
		& 7 x 32B at 128 & 1 x 128B at 128 & 1 x 128B at 128\\
		& 8 x 32B at 160 & 1 x 64B at 192 & 1 x 128B at 256\\
		& 8 x 32B at 192 & 1 x 32B 256 &\\
		& 8 x 32B at 224 & &\\
		& 1 x 32B at 256 & &\\ \hline
	\end{tabular}
\end{table}

Overall it is important when the data structures are designed to consider these basic metrics about the memory subsystem of the GPUs, so no restrictions will be hit and the memory will work with maximum throughput.

\subsection{CUDA algorithm}\label{cudaalg}
The CUDA algorithm uses the same parallel heuristics, that was introduced in Section \ref{heuristics}. The algorithm (Figure \ref{algorithm2}) itself follows the same execution path as the CPU based implementation. Thanks to the characteristics of the GPUs they can run with much more threads in parallel compared to the CPUs, thus doing more work at the same time.

\begin{figure}[!t]
	\centering
	\begin{algorithmic}[1]
		\Procedure {Parallel Community Detection CUDA}{$G(E,V,\omega)$}
		\State $Status\leftarrow initStatus(G(E,V,\omega))$
		\State $Parallel Louvain(G(E,V,\omega), Status)$
		\State $mod_{curr}\leftarrow Compute\ modularity$
		\State $newGraph\leftarrow Compute\ new\ input\ graph$
		\While {true}
		\State $Status\leftarrow initStatus(newGraph)$
		\State $Parallel\ Louvain(newGraph, Status)$
		\State $mod_{new}\leftarrow Compute\ modularity$
		\If {$mod_{new}-mod_{curr}<\Theta$}
		\State break
		\EndIf
		\State $mod_{curr}\leftarrow mod_{new}$
		\State $newGraph\leftarrow Compute\ new\ input\ graph$
		\EndWhile
		\EndProcedure
	\end{algorithmic}
	\caption{The community detection algorithm on CUDA. Input is graph $G(V,E,\omega)$. $\Theta$ is a predefined threshold}
	\label{algorithm2}
	
\end{figure}

\section{The parallel Louvain algorithm}\label{paralg}
The parallel algorithm (Figure \ref{algorithm1}) has the following major steps:

(1) Phases: Execute phases one at a time. Within each phase, multiple iterations are running and each iteration performs a parallel evaluation of the vertices without any locking mechanism using only the community information from the previous iteration. This is going on until the modularity gain between the iterations is not below the threshold.

(2) Graph rebuilding: After a successive phase the new community assignment is used to construct a new input graph for the next phase. This is done by introducing the communities in the new graph as vertices and the edges are added based on their connection to these communities.

\begin{figure}[!t]
	\centering
	\begin{algorithmic}[1]
		\Procedure {Parallel Louvain}{$G(E,V,\omega)$, Status}
		\State $Q_{curr}\leftarrow 0$
		\State $Q_{prev}\leftarrow -\infty$
		\While {true}
		\For {each $i\in V_k$ in parallel}
		\State $N_i\leftarrow Status.nodesToCom[i]$
		\For {each $j\in \Gamma(i)$}
		\State $N_i\cup {Status.nodesToCom[j]}$
		\EndFor
		\State $target\leftarrow arg \max_{t\in N_i} \Delta Q_{i\rightarrow t}$
		\If {$\Delta Q_{i\rightarrow target}>0$}
		\State $Status.nodesToCom[i]\leftarrow target$
		\EndIf
		\EndFor
		\State $Q_{curr}\leftarrow$ Compute new modularity
		\If {$|\frac{Q_{curr}-Q_{prev}}{Q_{prev}}|<\theta$}
		\State break
		\Else
		\State $Q_{prev}\leftarrow Q_{curr}$
		\EndIf
		\EndWhile
		\EndProcedure
	\end{algorithmic}
	\caption{The parallel Louvain algorithm for a single phase, inputs are the graph $G(V,E,\omega)$ and a Status object containing the initial community informations. $\theta$ is a user specified threshold}
	\label{algorithm1}
	
\end{figure}

\subsection{Implementation}\label{impl}
The implementation focuses on optimizations to provide maximum possible parallelization of the algorithm. The code is written in C++, for parallel primitives the NVIDIA provided Thrust library is used. Storing some intermediate values on the GPU is handled by the CUDPP hash tables. In this subsection the used data structures and procedures are going to be detailed.

\subsubsection{Data Structures}\label{data}
To generate the dendogram, the algorithm needs to know the nodes and edges building up the input graph. Looking at it in more detail it will become clear, that not all nodes need to evaluated during the computation. Specifically those nodes, which degree is $0$ will not be able to move to any other community, than their default initialized group. Because of this the computation will use only the edges as their source, target pairs will identify all nodes with degrees greater than $0$. All values required by the processings are stored in a {\em StatusCUDA} object. The input graph of each dendogram level is stored in {\em CUDAGraph} object.

To allocate device memory efficiently, {\em integer}, {\em unsigned integer} and {\em float} values are requested from the runtime driver as $3$ blocks. The memory pointers for individual variables are calculated on the host after allocation is done. To save up memory on intermediate storage variables the following fields are using the same memory space per block, as they do not interfere with each other during the computation.

Integer:
\begin{itemize}
	\item temp\_source $<->$ neighbourCommunities
	\item CUDAGraph.edgeSource\_temp $<->$ StatusCUDA.neighbourSource\_local
	\item CUDAGraph.edgeSource\_temp $<->$ StatusCUDA.neighbourSource\_local
\end{itemize}

Unsigned integer:
\begin{itemize}
	\item hash\_idx\_target $<->$ hash\_idx
	\item CUDAGraph.node $<->$ StatusCUDA.node
\end{itemize}

Float:
\begin{itemize}
	\item CUDAGraph.edgeWeight\_temp $<->$ StatusCUDA.neighbourWeights\_local
\end{itemize}

\subsubsection{Processes}\label{procs}
Computation of the dendogram involves the following processes: calculating neighbors of the input graph, initialization for the actual level, generation of new community assignments in the current level, renumbering the nodes based on the new community numbers, generating new input graph for the next level.

\paragraph{Neighbor computation}\label{neighbor}
As Subsection \ref{data} shows, the neighbors are stored in source-target arrays according to the nodes connected to each edge. The computation starts by first collecting all the neighbors. To achieve maximum parallelization on the neighbors, target-source pairs are copied at the end of the source-target pairs. The reason behind this is the undirected nature of the graph, nodes can move to a new community in each direction on the edge. There is no restriction on the order of the edges, which means at this stage the neighbors of a node may not be stored consecutively. This leads to ill-formed memory access patterns, that reduce the achievable memory bandwidth of the GPU. To fix this, the edges are sorted based on the source-target pairs using the {\bf thrust} library's {\em sort\_by\_key function} with the edge weight as value. Thanks to this sorting now the number of neighbors can be calculated. The {\em reduce\_by\_key} function will give back for all {\em neighbourSource} values the number of targets. To use this the algorithm will also need to know the index where the neighbors of a given node starts. By calling the {\em exclusive\_scan} function on the neighbor numbers, the result will give back the starting index of the targets for all given source.

\paragraph{Initializing status}\label{initstatus}
For every level of the dendogram, the nodes are assigned to their default cluster, that will be a group made from one node each. The variables storing intermediate results from the previous cycle are also initialized, also the degree for each node is calculated with the possible loops in the graph. During the initial step the {\em hash\_idx} variable is set to contain the node's index in the storing array. This is used to store the initial positions of the nodes in the hash table assigned to the {\em hash\_table\_handle} handler. For the degree computation it is necessary to know the index of the actual node where the result will be stored. As the degree computation will go through the weights for all neighbors, the first neighbor is taken and the source of that link is used to get the index from the hash table. This index helps to set the degree to the correct node.

\paragraph{Generating one level}\label{onelevel}
Each level in the dendogram contains those communities for each node where it will find a better modularity. This generation is iterative as it will move the nodes to new communities until the resulting gain on modularity is bigger, than a predefined threshold. Before the iterations, the variables (comSize, comSizeUpd, degreeUpd, internalUpd) used for updating the communities are initialized. {\em ComSize} will be always set to $1$, as at this point each node represents their own communities. {\em ComSizeUpd, degreeUpd and internalUpd} are all set to $0$ and are containing the change for a given community's size, degree and internal degree respectively. Also before the communities are processed the actual state of the {\em neighbourWeights} variable is copied into {\em neighbourWeights\_store}.\\

In every iteration first the indexes of the individual target values of each edge are taken from the {\em hash\_table\_handle} stored hash table into the {\em hash\_idx} variable. The next step is to compute the actual community of the given targets. This is important as the computation will work on communities and they can change in every iteration for all the edges. This is done by the {\em neighbourCommunitySetKernel} device kernel, that takes the current {\em nodesToCom} values for the actual communities based on the indexes stored in the {\em hash\_idx}. Also for those edges, that represents loops the weight is getting set to $0$. This is done because the computation would like to find a better community among a given node's neighbors. If one node will belong to a false community, but with many different nodes there is a good chance it will never be able to find the correct cluster. Even if the original graph didn't have loops, the computation might lead to clusters, where multiple edges will stay within the same community.\\

After having the communities first a reordering is required. Using {\em sort\_by\_key} with {\em neighbourSource} and {\em neighbourCommunities} as keys the {\em neighbourWeights} values are reordered to have the edges in a consecutive order based on the communities. This is used to calculate the weight of all the given communities with the {\em reduce\_by\_key} function. The reduction's input keys will be the same as for the sorting, with {\em neighbourSource\_local} and {\em neighbourCommunities\_local} being the result keys and {\em neighbourWeights\_local} storing the final reduced values.

In the final stage the number of neighbors and their starting coordinates are calculated as it was detailed before for the neighbor computation.\\

By having the current communities set and their values calculated the next move of the nodes can be processed. The {\em SetNewCommunityWarpShuffle} device kernel is responsible for calculating each node's next community. To make this efficiently every CUDA block will compute one node. The threads within the block will get one neighboring community from the list of all clusters if the node has any neighbors. The first thread will fetch the current community for the node being evaluated. What needs to be done is effectively a maximum search based on multiple criteria. The neighbor providing the best weight increase will be selected. If the community of this selected neighbor is the same where the processed node is, nothing happens. If multiple neighbors are providing the same increase, then the community with the lower index will be selected based on what is detailed in Subsection \ref{genmin}.

As the name of the kernel implies, the implementation uses warp shuffle operations. With this specific values can be shared among the threads within a thread block without requiring any addition global memory read or write operations, further increasing the access performance of the algorithm.\\

As in the current iteration the neighbors will not be used anymore the weight difference and the selected community will be stored in the node's first neighbor's weight and community value respectively.\\

The changes will be registered by the {\em oneLevelKernel} device kernel. If the computed next community is different than the one already occupied by the node the {\em remove} device function will remove the weight from the old community associated with the node and the {\em insert} device function will add it to the new cluster. Additionally {\em comSizeUpd} is changed as needed with {\em atomicSub} and {\em atomicAdd} to avoid issues from multiple concurrent updates. Finally the {\em nodesToComNext} for each node will contain the community it will belong to in the next iteration.\\

After the computations are done the modularity is recomputed to check if additional iterations should be taken. If yes, the {\em communitySetKernel} device kernel will apply the changes on each cluster, that was computed throughout the iteration. To prepare for the next pass the weights calculated for the graph are copied back from {\em neighbourWeights\_store} into {\em neighbourWeights}. Finally the {\em nodesToComPrev} is set to {\em nodesToCom}, {\em em nodesToCom} to {\em nodesToComNext} and {\em nodesToComNext} will be what was computed in the previous step.

When all possible iterations are done one final step is applied on those nodes, that have only one neighbor community. On those the {\em mergeInitKernel, mergeIsolatedNodeKernel, mergeSetKernel} device kernels are pushing them into those neighbors to increase quality of the end result. Finally the modularity is recomputed for later decision if multiple levels of the dendogram are necessary or not.

\paragraph{Renumbering nodes}\label{renumber}
Finishing the generation of a new level in the dendogram might lead to clusters getting removed as all their nodes are moved to other communities. To keep the data organized, each level of the dendogram will have consecutive cluster numbering, which means after one level is computed, the IDs of the groups will be mapped into a continuous array.\\

To do this, first the {\em nodesToCom} array containing the actual cluster assignments is copied into {\em nodesToComPrev} for preservation. The values in {\em nodesToComPrev} needs to be ordered, so {\em sort\_by\_key} will sort the cluster IDs and for the values {\em node} and {\em hash\_idx} is used, where {\em hash\_idx} will contain the indexes of the nodes. The resulting {\em nodesToComPrev} will be first copied into {\em nodesToCom} to have a copy of the sorted values.

Following this using the {\em unique} function from {\bf thrust} on {\em nodesToComPrev} and {\em hash\_idx} as keys, only the different cluster IDs will remain with their indexes. Now only the {\em nodesToComPrev} needs to be ordered based on {\em hash\_idx} as key, resulting in the unique clusters, ordered by their indexes, which points to their location in {\em nodesToCom}.\\

After computing the unique clusters, they will be pushed with their indexes into a hash table assigned to the {\em commMap\_hash\_table\_handle} handler. Now from this table the indexes for the clusters stored in {\em nodesToCom} can be retrieved. Finally this index will be the new community for the respective nodes, so at this point {\em hash\_idx} is the community and {\em node} is the node that needs to be transfered to the host, so each level of the dendogram can be stored on the host. This node, community pair will also be stored in another hash table assigned to {\em nodeCommMap\_hash\_table\_handle}, that will be used in the last process.

\paragraph{Inducing new graph}\label{inducedgraph}
Each level of the dendogram will need an input graph to compute on. At the first level this will be the original graph, but for the subsequent levels this graph will be generated from new cluster assignments. At the end of the renumber process, each node has a new ID associated to them, this will represent the new cluster ID in the next level's graph.

First the algorithm has to compute the unique IDs, so it will know how many different communities there are, which will tell how many nodes the graph in the next level will have. As these IDs are already stored in a hash table, the original storage can be manipulated. At the beginning the {\em hash\_idx} array will be sorted. After that the {\em unique\_copy} function will collect and copy the unique clusters into {\em cuGraph.node}. This way the nodes for the next level are generated.\\

To compute the new edges, again it is used, that the links are stored in source, target pairs. First the indexes of the {\em neighbourSource} values are retrieved into the {\em hash\_idx\_source} array from the {\em nodeCommMap\_hash\_table\_handle} pointed hash table. Then the same is done for the targets, in that case the indexes will be stored in {\em hash\_idx\_target}. Because the original graph will not be used again, the new edges will be computed at the same memory address. The two {\em hash\_idx} values retrieved before will give the new source and target IDs for the edges.

To finish the computation only the weights are missing at this point. As it was detailed in Section \ref{prob}, the graph can be unweighted, in which case the weights are considered to be $0$ for all links. In every case where a given weight is $0$, the algorithm will always increase that value, so during computation all edges have a weight greater than $0$. Here a {\em neighbourSource, neighbourTarget} IDs as keys will sort the {\em neighbourWeights} values. This way a call to {\em reduce\_by\_key} on these variables will result in the new edges, that are stored in {\em edgeSource\_temp, edgeTarget\_temp} and {\em edgeWeight\_temp}. Copying these values back to {\em neighbourSource, neighbourTarget} and {\em neighbourWeights} respectively will conclude the computation of one level.

\section{Evaluation}\label{results}
\begin{table}[!th]
	\caption{The test data}
	\label{test}
	\centering
	\begin{tabular}{|c|c|c|c|c|c|}
		\hline
		Input graph & Num. vertices (n) & Num. edges (m) & Max. deg. & Avg. deg. & RSD\\
		\hline
		CNR & 325557 & 2738970 & 18236 & 16,826 & 13,024\\
		coPapersDBLP & 540486 & 15245729 & 3299 & 56,414 & 1,174\\
		Channel & 4802000 & 42681372 & 18 & 17,776 & 0,061\\
		Europe-osm & 50912018 & 54054660 & 13 & 2,123 & 0,225\\
		Soc-LiveJournal1 & 4847571 & 68475391 & 22887 & 28,251 & 2,553\\
		MG1 & 1280000 & 102268735 & 148155 & 159,794 & 2,311\\
		Rgg\_n\_2\_24\_s0 & 16777216 & 132557200 & 40 & 15,802 & 0,251\\
		uk-2002 & 18520486 & 261787258 & 194955 & 28,270 & 5,124\\
		NLPKKT240 & 27993600 & 373239376 & 27 & 26,666 & 0,083\\
		MG2 & 11005829 & 674142381 & 5466 & 122,506 & 2,370\\
		friendster & 51952104 & 1801014245 & 8603554 & 69,333 & 17,354\\
		\hline
	\end{tabular}
\end{table}

Testing was done on multiple real world networks (Table \ref{test}), from which the first two was used for experimental study, the remaining graphs are for theoretical evaluation. The limit was the available resources, as only the first two graph could fit in the system's memory. In this section the runtime of the parallel algorithm will be compared between the CPU and GPU implementation.

Evaluation was done using the following (Table \ref{system}) system:

\begin{table}[!th]
	\caption{The test system}
	\label{system}
	\centering
	\begin{tabular}{|c|c|c|c|c|}
		\hline
		CPU & GPU & OS & Compiler & CUDA\\
		\hline
		Intel Core i7 & GeForce GTX & Ubuntu & GCC & 9.0\\
		4710HQ & 980 & 16.04 & 4.8.5 & \\
		\hline
	\end{tabular}
\end{table}

On the CPU, parallelization was running on $threads_{cpu}=8$ threads. As the GTX 980 has $16$ {\em Multiprocessors} and each of them can have maximum $2048$ threads, in overall giving $threads_{gpu}=32768$ threads, that can run in parallel at the same time.

\subsection{Experimental results}\label{runtime1}
Experimental results as mentioned before, are provided for the CNR and coPapersDBLP graphs. The whole community detection has multiple steps as it was described in Section \ref{louvain}. 

\begin{figure}[!th]
	\centering
	\includegraphics[width=\linewidth]{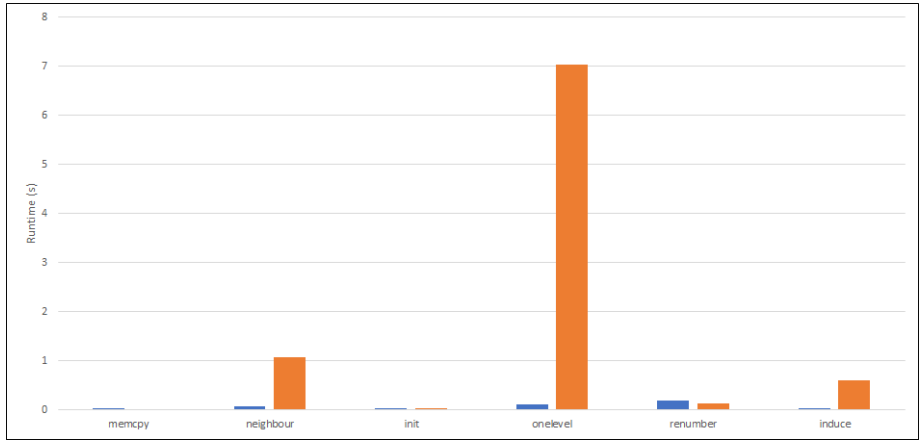}
	\caption{Detailed runtime of the CNR graph}
	\label{gpu_cnr}
\end{figure}

The overall computation time on the CPU is $t_{cpu}=9,05 s$ and on the GPU it is $t_{gpu}=0,444 s$. Detailed runtimes are presented on Figure \ref{gpu_cnr}.

\begin{figure}[!th]
	\centering
	\includegraphics[width=\linewidth]{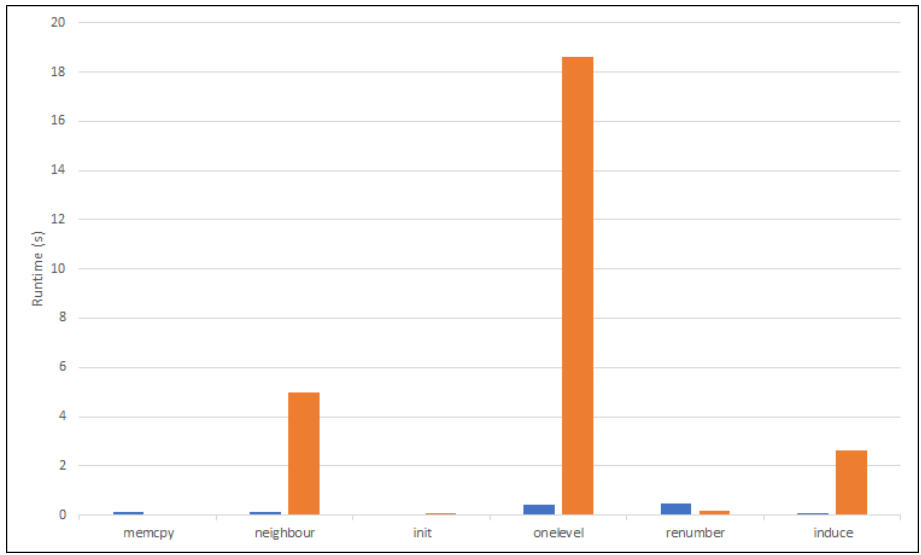}
	\caption{Detailed runtime of the coPapersDBLP graph}
	\label{gpu_dblp}
\end{figure}

Brake down on the different processes runtime is presented on Figure \ref{gpu_dblp}.

Here the overall computation time on the CPU is $t_{cpu}=26,795 s$ and on the GPU it's $t_{gpu}=1,188 s$.

\subsection{Theoretical results}\label{runtime2}
The problem with the following graphs is that their size extend beyond the available system resources, but the average degree of nodes is known for every network, so it was used together with the number of nodes and edges to compute the theoretical performance (Table \ref{theor} and Table \ref{theorCPU}) of the parallel implementations. These runtimes for neighbour calculation and to generate one dendogram level, are calculated by dividing the execution time of one of the previous graphs (let's take CNR) with the mentioned average degree. Finally this number is multiplied with the average degree of the bigger datasets. The runtime for the rest of the processes is calculated using the number of nodes and edges, so the execution time of init and renumber will be computed by dividing with the node count and then multiplying with the bigger networks vertex count, while for the induce graph function the same is done, but with the edge count. Have to mention that when dividing to compute the runtime for one unit, also has to multiple with $threads_{gpu}$, because during computation one unit will hold the maximum threads running in parallel, thus the runtime needs to be transformed back, like how it will be in the sequential computation. On multiplications on the other hand the result needs to be divided by this thread number to get the final result.

\begin{table}[!th]
	\label{theor}
	\centering
	\begin{tabular}{|c|c|c|c|c|c||c|}
		\hline
		Input graph & Neighbour & Init & Onelevel & Renumber & Induce & Full\\
		\hline
		Channel & 0,067 & 14,38 & 0,11 & 2,8 & 0,55 & 3,88\\
		Europe-osm & 0,008 & 152,47 & 0,05 & 29,71 & 0,69 & 34,21\\
		Soc-LiveJournal1 & 0,11 & 14,52 & 0,19 & 2,83 &	0,88 & 4,36\\
		MG1 & 0,61 & 3,83 &	0,48 & 0,75 & 1,31 & 3,24\\
		Rgg\_n\_2\_24\_s0 & 0,06 & 50,25 & 0,18 & 9,79 & 1,69 & 12,96\\
		uk-2002 & 0,11 & 55,47 & 0,81 & 10,81 & 3,35 & 16,43\\
		NLPKKT240 & 0,1 & 83,84 & 0,28 & 16,34 & 4,77 & 23,55\\
		MG2 & 0,47 & 32,96 & 5,86 & 6,42 & 8,61 & 22,18\\
		friendster & 0,26 & 155,59 & 1,16 & 30,32 & 23,01 & 58,59\\
		\hline
	\end{tabular}
	\caption{Theoretical runtimes on the GPU}
\end{table}

\begin{table}[!th]
	\label{theorCPU}
	\centering
	\begin{tabular}{|c|c|c|c|c|c||c|}
		\hline
		Input graph & Neighbour & Init & Onelevel & Renumber & Induce & Full\\
		\hline
		Channel & 1,14 & 0,37 & 7,43 & 1,93 & 9,24 & 20,11\\
		Europe-osm & 0,14 & 3,91 & 0,89 & 20,49 & 11,7 & 37,12\\
		Soc-LiveJournal1 & 1,81 & 0,37 & 11,82 & 1,95 & 14,83 & 30,77\\
		MG1 & 10,21 & 0,09 & 66,83 & 0,52 & 22,14 &	99,79\\
		Rgg\_n\_2\_24\_s0 & 1,01 & 1,29 & 6,61 & 6,75 & 28,69 & 44,36\\
		uk-2002 & 1,81 & 1,42 & 11,82 & 7,45 & 56,68 & 79,18\\
		NLPKKT240 & 1,7 & 2,15 & 11,15 & 11,26 & 80,81 & 107,08\\
		MG2 & 7,83 & 0,85 & 51,23 & 4,43 & 145,96 & 210,29\\
		friendster & 4,43 & 3,99 & 28,99 & 20,9 & 389,93 & 448,25\\
		\hline
	\end{tabular}
	\caption{Theoretical runtimes on the CPU}
\end{table}

\subsection{Analysis}\label{analysis}
Starting with the experimental results, the detailed runtimes shows, that the computation of one phase as described in Algorithm \ref{algorithm1} takes most of the time running on the CPU.

For the "CNR" graph to generate one level of the dendogram the CPU takes $t_{onephase_{cpu}}=7,037 s$ to finish, on the GPU it is $t_{onephase_{gpu}}=0,975 s$. Computing the new graph input for the consecutive phase on the CPU is $t_{induce_{cpu}}=0,593 s$ and on the GPU is $t_{induce_{gpu}}=0,035 s$. Processing the neighbours on the CPU takes $t_{neighbour_{cpu}}=1,075 s$, while the GPU requires $t_{neighbour_{gpu}}=0,064 s$. The GPU  is also hit with an additional memory transfer time as the input and output values have to be copied between the host and device. This is $t_{memcpy}=0,024 s$. Overall the GPU is $~20$ times faster than the CPU implementation.

The runtimes for the "coPapersDLBP" graph are the following: generating the dendogram's level on the CPU takes $t_{onephase_{cpu}}=18,59 s$, on the GPU it is $t_{onephase_{gpu}}=0,4 s$. Computation of the graph induction the CPU runs for $t_{induce_{cpu}}=2,62 s$ and on the GPU for $t_{induce_{gpu}}=0,007 s$. Computing the neighbours on the CPU takes $t_{neighbour_{cpu}}=4,97 s$, while on the GPU it's $t_{neighbour_{gpu}}=0,11 s$. The overall time of the device memory copy operations is $t_{memcpy}=0.11 s$. Overall the GPU achieves $~23$ times better performance.

The theoretical results also shows similar performances, the speedups for the networks respectively are: $5$, $N/A$, $7$, $31$, $3$, $5$, $5$, $9$, $8$.
For the Europe-osm graph the $N/A$ shows, that there was no speed up, actually the GPU was running a few percent slower, than the CPU counter part. The GPU based theoretical runtimes (Table \ref{theor}) show, that for this graph the renumber part is considerably slower, than for any other graphs. While this encourages further evaluation of the process in the whole clustering, it already shows that not just the size of the graph is the deciding factor of the performance, but the structure of the network as well.

One aspect of the algorithm that can reduce the efficiency of the GPU is the diverging execution path on the different threads, thanks to the different number of neighbors of the nodes. This might be further optimized by for example reordering the nodes to have the nodes with equal or nearly as much number of neighbors being consecutive in memory \cite{msc}.

\section{Future work}\label{future}
Implementing the other heuristics from \cite{parlouvain} might further increase the performance of the GPU implementation. Optimization should be explored on the memory management level to make the algorithm be able to process bigger graphs on the given system. As seen in section \ref{analysis}, while the GPU version greatly outperforms the CPU, some phases are much slower on the CPU. These should be further explored to reduce the required time to compute these processes too.

\section{Conclusion}
By implementing the heuristics described in Section \ref{heuristics} using the CUDA platform (Section \ref{CUDA}) for the GPU version (Section \ref{cudaalg}), the experimental results show a $12$ times faster computation time, while the theoretical study points up to a $31$ times speedup over the CPU based parallel version of the Louvain algorithm. According to the results detailed in Section \ref{analysis}, it can be said that the GPU's can be used more effectively if the underlying data is big enough, but is not the only important factor. In the case of this algorithm, the hierarchy of the data is also significant in achieving higher performances. The test cases show, that the time to generate the new level of the dendogram was always significantly lower compared to the time of generating the new graph for the next pass: $t_{onephase_{gpu}}<<t_{induce_{gpu}}$. The CPU was also producing similar results: $t_{onephase_{cpu}}<t_{induce_{cpu}}$. The runtimes are leading to the conclusion, that the parallel Louvain modularity (Algorithm \ref{algorithm1}) can profit by running on the GPUs as $t_{onephase_{gpu}} << t_{onephase_{cpu}}$ is valid in all the test cases and also the same applies to the other computational phases as $t_{neighbor_{gpu}} << t_{neighbor_{cpu}}$ and $t_{induce_{gpu}} << t_{induce_{cpu}}$ all stands.

\section{Acknowledgments}
The GPU used for the computations was provided by the Wigner RCP's GPU Laboratory. This research did not receive any specific grant from funding agencies in the public, commercial, or not-for-profit sectors. A preliminary version of this paper appeared in \cite{ines}.

\section*{References}

\bibliography{mybibfile}

\end{document}